\documentclass[11pt,twoside]{article}  
\usepackage{adassconf}
\usepackage{epsfig}

\begin{document}   

\paperID{P7-1}

\title{Automated Determination of Stellar Population Parameters in
Galaxies Using Active
Instance-based Learning }
\titlemark{Automated Determination of Stellar Population Parameters }
\author{Thamar Solorio, Olac Fuentes, Roberto Terlevich, Elena Terlevich}
\affil{INAOE, Luis Enrique Erro \#1, Tonantzintla, Puebla, 72840,
M\'{e}xico}
\author{Alessandro Bressan}
\affil{Osservatorio Astronomico di Padova Vicolo dell'
Osservatorio 5, 1-35122 Padova, Italy}

\contact{Thamar Solorio} \email{thamy@ccc.inaoep.mx}

\paindex{Solorio, T.} \aindex{Fuentes, O.} \aindex{Terlevich, R.}
\aindex{Terlevich, E.} \aindex{Bressan, A.}

\authormark{Solorio \& Fuentes \& Terlevich \& Terlevich \& Bressan}


\begin{abstract}          
In this work we focus on the determination of the relative
distributions of young, intermediate-age and old populations of
stars in galaxies. Starting from a grid of theoretical population
synthesis models we constructed a set of  model
galaxies with a distribution of ages, metallicities and intrinsic
reddening. Using this set we have explored a new fitting method
that presents several advantages over conventional methods. We
propose an optimization technique that combines active learning
with an instance-based machine learning algorithm. Experimental
results show that this method can estimate with high speed and
accuracy the physical parameters of the stellar populations.
\end{abstract}

\section{Introduction}

The availability for the first time of huge astronomical
spectroscopic surveys such as the SDSS, with more than
$10^{6}$ spectra, will allow the determination of
intrinsic physical parameters of a large number of galaxies,
including the age distribution or star formation history and
metallicity distribution of their stellar populations.

The importance of the accurate knowledge of these parameters for
cosmological studies and for the understanding of galaxy formation
and evolution cannot be overestimated. Template fitting has been
used to carry out estimates of the distribution of age and
metallicity from spectral data. Although this technique achieves
good results, it is very expensive in terms of computing time and
therefore can be applied only to small samples.

Starting from a grid of theoretical population synthesis models we
constructed a set of model galaxies with a
distribution of ages, metallicities and intrinsic reddening. Using
this set we have explored a new method that maximizes speed and
accuracy. Our proposed technique combines standard least-squares
fitting with an active instance-based machine learning algorithm.
Experimental results show that this method can estimate with high
speed and accuracy the physical parameters of the stellar
populations.  Based on empirical evidence we believe that this
method can be applied with equal success to other astronomical
problems, reducing the computational cost and thus providing the
capability of analyzing larger quantities of astronomical data.
\section{Description of the Models}
For the spectral synthesis of simple stellar populations the
atmospheric models have been folded with the predicted number of
stars along isochrones of given age and metal content (Bressan et
al. 1994). The atmosphere models have been inserted in low
resolution Kurucz models (Kurucz 1993) in order to preserve the
complete energy distribution.

The models have the following characteristics:
\begin{itemize}
\item Ages are from $10^{6}yr$ to $2\times 10^{10}yr$ in
logarithmic steps: \newline
[$10^{6}yr,10^{8}yr$,$10^{8.3}yr$,$10^{8.6}yr$,$10^{9}yr$,$10^{9.6}yr$,$10^{9.78}yr$,$10^{10}yr$
,$10^{10.2}yr$]
\item Metallicity has the values Z=[0.0004, 0.004, 0.008, 0.02,
0.05] in Solar units \item The resolution is smoothed at the desired
value.
\end{itemize} For the present experiments we used solar metallicity (0.02)
and a resolution of 20 \AA.

\section{The Proposed Solution}
Given an observed galaxy spectrum we would like to determine the
relative distribution of ages and their intrinsic
reddening. We restricted the problem to finding three
contribution of ages: starbursts of age 1Myr, an intermediate age
population with age between 100Myr and 1000Myr and an old population
with age greater than 1000Myr. Each of the three populations is
affected by the same reddening law which is defined as follows:
\begin{equation}\label{reddening}
R(c_{i},\lambda )=1-e^{\lambda \times c_{i}}
\end{equation}
where $c_{i}$ is the free parameter of each stellar population and
$\lambda$ is the wavelength, in this case going from 890\AA
\hspace{.01in} to 2.301 $\mu$m.
In order to determine the free
parameters of reddening and the relative contributions we pose the
problem as an optimization problem, where a modified version of a
machine learning algorithm is trained to estimate the reddening
parameters of the three populations. Once we have an estimate of
the reddening we can compute the relative contribution of ages,
$\overrightarrow{A}$, with a pseudo inverse matrix as follows:
\newline Let $M=[\overrightarrow{m_{1}},...,\overrightarrow{m_{9}}]$ be
the grid of our nine
theoretical models
described earlier. $\overrightarrow{o}$ is the observed spectrum,
and $\overrightarrow{r}=c_{1},c_{2},c_{3}$ is the vector of the
free reddening parameters predicted by the learning algorithm for
$\overrightarrow{o}$. We can compute
$S=[\overrightarrow{F_{1r}},...,\overrightarrow{F_{9r}}]$, by
applying to the theoretical models the reddening function as
defined in equation~\ref{redSpectra}.
\begin{equation}\label{redSpectra}
   \overrightarrow{F_{ir}}(\lambda )=\overrightarrow{m_{i}}(\lambda )
\times R(c_{i},\lambda )
\end{equation}
We know that the observed spectrum $\overrightarrow{o}$ is the
product of $S$ and the unknown relative contributions
$\overrightarrow{A}$,
\begin{equation}\label{testSpec}
\overrightarrow{o}=\overrightarrow{A} \times S
\end{equation}
\begin{equation}\label{abund}
    \overrightarrow{A}=S^{*}\times \overrightarrow{o}
\end{equation}
then by computing $S^{*}$, the pseudo-inverse of $S$, we can
determine the relative contribution of ages, as equation
\ref{abund} shows. The following section introduces the
optimization procedure used in this work.
\section{The Optimization Procedure}
We are interested in the problem of finding the parameters of a
known analytic function that best match an observation. Let
$\overrightarrow{o}$ be the observed galactic spectrum variable,
let $f(\overrightarrow{r})$ be a function with the same
dimensionality as $o$. The goal of the optimization procedure is
to obtain the value of $f(\overrightarrow{r})$ that minimizes the
error $e=|o- f(\overrightarrow{r})|$. In order to solve the
problem more efficiently, we pose it as a learning problem, where
a learning algorithm learns the reddening parameters
\overrightarrow{r}, and with a forward model we compute
$f(\overrightarrow{r})$. The training set used by the algorithm,
$\langle f(\overrightarrow{x_{i}}),\overrightarrow{t}_{i}\rangle
$, is formed by randomly generated reddening parameters,
$\overrightarrow{x_{i}}$, and their corresponding galactic
spectra, \overrightarrow{t_{i}}, where contributions of ages were
also generated randomly; its test set consists of the galactic
spectra to be analyzed denoted here by
$\overrightarrow{o}_{1},...,\overrightarrow{o}_{n}$ and it outputs
an estimate of $\overrightarrow{r_{1}},...,\overrightarrow{r_{n}}$
that is expected to minimize the errors $e_{1},...,e_{n}$. When a
new set of solutions
$\overrightarrow{r_{1}},...,\overrightarrow{r_{n}}$ is proposed by
the algorithm, we compute their corresponding
$f(\overrightarrow{r_{1}}),...,f(\overrightarrow{r_{n}})$, using
equations \ref{redSpectra},\ref{testSpec} and \ref{abund}, and use
the new pairs $\langle
f(\overrightarrow{r_{i}}),\overrightarrow{o}'_{i}\rangle $ to
augment the training set, and continue this iterative process
until convergence is attained. Since this type of active learning
adds to the training set examples that are progressively closer to
the points of interest, the errors are guaranteed to decrease in
every iteration. The pseudocode of the algorithm is the following:
        \begin{small}
        \begin{enumerate}
            \item Generate randomly an initial set of vectors
$\overrightarrow{x}_{i},...\overrightarrow{x}_{m}$ and compute their
corresponding
$f(\overrightarrow{x}_{1}),..., f(\overrightarrow{x}_{m})$.
            \item Let $P = {\langle
f(\overrightarrow{x}_{1}),\overrightarrow{t}_{1}\rangle,
...,\langle f(\overrightarrow{x}_{m}),\overrightarrow{t}_{m}\rangle }$ be
the initial training
set.
            \item Let $T ={\overrightarrow{o}_{1},...,
\overrightarrow{o}_{n}}$ be the test
set.
            \item While $T$ is not empty
                  \begin{enumerate}
                  \item Train an approximator A using P as training set
                  \item For each  $\overrightarrow{o}_{i}$ in $T$
                  \begin{itemize}
                  \item Use A to predict $\overrightarrow{r}'_{i}$
                  \item Generate $\overrightarrow{o}'_{i}$
                  \item $P = P \cup \langle f(\overrightarrow{r}'_{i}),
\overrightarrow{o}'_{i}\rangle$
                  \item If $|o_{i} -\overrightarrow{o}'_{i}| <threshold$
remove $o_{i}$ from
                  $T$
                  \end{itemize}
                  \end{enumerate}
        \end{enumerate}
        \end{small}
In this problem the approximator mentioned in step 4.1 is Locally
Weighted Linear Regression, an instance-based learning algorithm
that has shown good results in similar optimization problems
(Fuentes \& Solorio 2003).

\section{Experimental Results}

\begin{figure}[t]
\centering
\begin{small}
\begin{tabular}{c}
  \includegraphics[height=45mm, width=95mm]{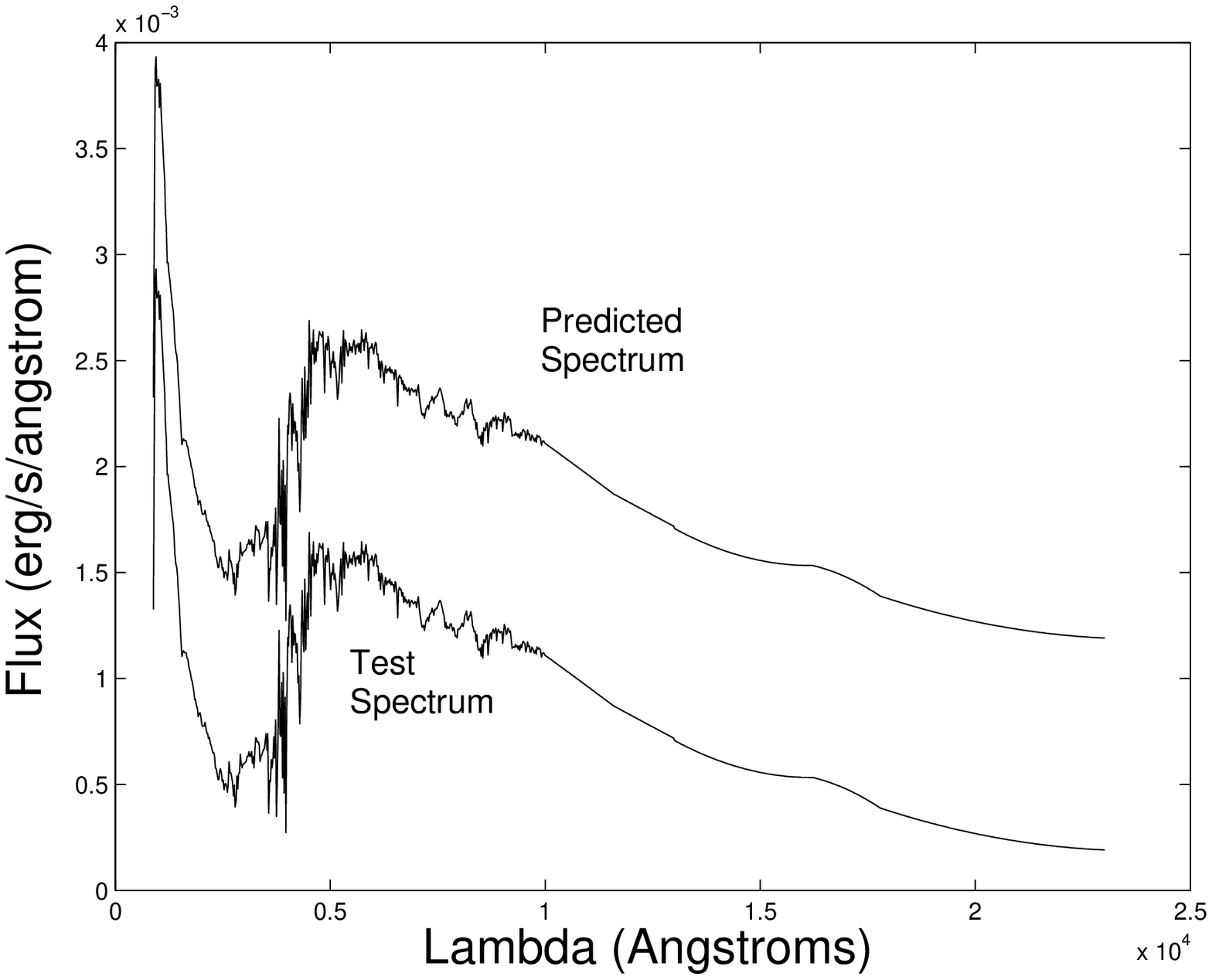}\\
 \small A=[0.1132, 0.1763, 0, 0, 0, 0, 0, 0, 0.7105],R=[-0.0004, -0.0011,
-0.0008]\\
\small A'=[0.1132, 0.1763, 0, 0, 0, 0, 0, 0, 0.7105],R'=[-0.0005, -0.0011,
-0.0008]\\
 \end{tabular}
 \caption{\small{In this figure we show test and predicted spectra shifted
by a constant amount
to aid visualization. Vectors A and R are the parameters for the test
spectrum, while A' and R'
are the corresponding predicted parameters.}} \label{f:exam}
\end{small}
\end{figure}
\begin{table}
\centering
\begin{tabular}{|c|c|c|c|}
  \hline
   & \begin{small}$r_{1}$\end{small} & \begin{small}$r_{2}$\end{small} &
\begin{small}$r_{3}$\end{small} \\
  \hline
  \begin{small}mae$\times 10^{6}$\end{small} &
\begin{small}0.0149\end{small} &
\begin{small}0.0482\end{small}& \begin{small}0.4182\end{small}\\
  \hline
\end{tabular}
\caption{\small Mean absolute errors in reddening parameters}
\label{t:redpars}
\end{table}
\begin{small}
\begin{table}[t]
\centering
\begin{tabular}{|c|c|c|c|c|c|c|c|c|c|}
  \hline
   & \small $A_{1}$ & \small $A_{2}$ & \small $A_{3}$  & \small $A_{4}$  &
\small $A_{5}$  &
\small $A_{6}$  &\small $A_{7}$  & \small $A_{8}$  & \small $A_{9}$  \\
  \hline
  \begin{small}mae$\times 10^{6}$\end{small} & \small 4.58 & \small 2.92&
\small 1.79 & \small
2.78& \small 6.48& \small 2.83& \small 5.79& \small 4.33& \small1.90\\
  \hline
\end{tabular}
\caption{\small Mean absolute errors in predicted population fractions}
\label{t:abund}
\end{table}
\end{small}
In order to evaluate our proposed solution we experimented
generating randomly 500 spectra together with metallicities and
intrinsic reddening, we then generated their corresponding
spectra. From this set we selected randomly 150 spectra that were
used as the test set, the remaining spectra were used as the training set.
We repeated this process 10 times, and reported the overall
average. Table~\ref{t:abund} presents mean absolute errors in
estimating age distributions, in Table~\ref{t:redpars} we show the
errors in the reddening parameters. Figure~\ref{f:exam} shows a
comparison between a test example and the predicted one. On
average, it takes 15 seconds to predict the parameters of a single
spectrum.
\section{Conclusions}
We presented in this work an optimization algorithm that can
estimate with high accuracy age distributions and reddening of
stellar population in galaxies. The algorithm achieves convergence by
iteratively creating new data points that lie in the vicinity of
the query point. One important feature of this method is its high
speed, it takes 15 seconds to estimate the parameters of a single
spectrum. This represents a great advantage over other more
conventional methods proposed for this problem, which may take
several hours to find the solution for a single spectrum.



\begin{references}
\reference Bressan A.\ \& Chiosi C.\ \& Fagotto F.\ 1994, ApJS 94,63
\reference Fuentes O.\ \& Solorio T.\ 2003, AIA2003, Spain
\reference Kurucz R. L.\ 1993, CD-ROM13: Atlas9, SAO,Harvard, Cambridge
\end{references}
\end{document}